	\documentclass[twoside,magazine]{IEEEtran}
	\usepackage{makecell}
	\usepackage{hyperref}
	\usepackage{array}
	\usepackage{graphicx,amssymb,amsmath}
	\usepackage{multicol}
	\usepackage[noadjust]{cite}
	\usepackage{setspace}
	\usepackage{subfig}
	\usepackage{graphicx}
	\usepackage{float}
	\usepackage {url}
	\usepackage{stfloats}
	\usepackage{amsthm,pifont}
	\usepackage{flushend}
	\usepackage{cases,subeqnarray}
	\usepackage{bm,multirow,bigstrut}
	\usepackage{amsmath, amsthm, amssymb}
	\usepackage{textcomp}
	\usepackage{latexsym,bm}
	\usepackage{booktabs}
	\usepackage{xcolor}
	\usepackage{mathtools}
	\usepackage{dsfont}
	\usepackage{extarrows}
	\usepackage{epsfig}
	\usepackage{epsfig}
	\usepackage{epstopdf}
	\usepackage[noend]{algpseudocode}
	\usepackage{algorithmicx,algorithm}
	\usepackage{xspace}
	\usepackage{listings}
	\usepackage{xcolor}
	\definecolor{commentcolor}{RGB}{85,139,78}
	\definecolor{stringcolor}{RGB}{206,145,108}
	\definecolor{keywordcolor}{RGB}{0,0,128}
	\definecolor{backcolor}{RGB}{220,220,220}
	\theoremstyle{plain}

	\theoremstyle{plain}

	\usepackage{amsmath}

	\IEEEoverridecommandlockouts
	
\begin{document}
	\title{Exploring Attention-Aware Network Resource Allocation for Customized Metaverse Services}
	\author{Hongyang Du, Jiacheng Wang, Dusit~Niyato,~\IEEEmembership{Fellow,~IEEE}, Jiawen~Kang, Zehui Xiong, Xuemin~(Sherman)~Shen,~\IEEEmembership{Fellow,~IEEE}, and Dong~In~Kim,~\IEEEmembership{Fellow,~IEEE}
			\thanks{H.~Du, J.~Wang and D. Niyato are with the School of Computer Science and Engineering, Nanyang Technological University, Singapore (e-mail: hongyang001@e.ntu.edu.sg, jcwang\_cq@foxmail.com, dniyato@ntu.edu.sg).}
\thanks{J. Kang is with the School of Automation, Guangdong University of Technology, China. (e-mail: kavinkang@gdut.edu.cn)}
\thanks{Z. Xiong is with the Pillar of Information Systems Technology and Design, Singapore University of Technology and Design, Singapore (e-mail: zehui\_xiong@sutd.edu.sg)}
\thanks{X. Shen is with the Department of Electrical and Computer Engineering, University of Waterloo, Canada (e-mail: sshen@uwaterloo.ca)}
\thanks{D. I. Kim is with the Department of Electrical and Computer Engineering, Sungkyunkwan University, South Korea (e-mail: dikim@skku.ac.kr)}
	}
	\maketitle
	\vspace{-1cm}
	\begin{abstract}
Emerging with the support of computing and communications technologies, Metaverse is expected to bring users unprecedented service experiences. However, the increase in the number of Metaverse users places a heavy demand on network resources, especially for Metaverse services that are based on graphical extended reality and require rendering a plethora of virtual objects. To make efficient use of network resources and improve the Quality-of-Experience (QoE), we design an attention-aware network resource allocation scheme to achieve customized Metaverse services. The aim is to allocate more network resources to virtual objects in which users are more interested. We first discuss several key techniques related to Metaverse services, including QoE analysis, eye-tracking, and remote rendering. We then review existing datasets and propose the user-object-attention level (UOAL) dataset that contains the ground truth attention of $30$ users to $96$ objects in $1,000$ images. A tutorial on how to use UOAL is presented. With the help of UOAL, we propose an attention-aware network resource allocation algorithm that has two steps, i.e., attention prediction and QoE maximization. Specially, we provide an overview of the designs of two types of attention prediction methods, i.e., interest-aware and time-aware prediction. By using the predicted user-object-attention values, network resources such as the rendering capacity of edge devices can be allocated optimally to maximize the QoE. Finally, we propose promising research directions related to Metaverse services.

	
	\end{abstract}
	\IEEEpeerreviewmaketitle
	\section{Introduction}
 The concept of ``Metaverse" embodies our good expectations for the future of the Internet~\cite{tang2022roadmap}. Unlike current Internet, in which we need to browse content through screens, Metaverse can be considered as a virtual Internet that descends on the physical world: It is virtual, yet rooted in real-world data, and can interact with the physical world. In Metaverse, everyone can customize the digital world, simulate the real world, trade, and simply have fun. Thus, it is believed that Metaverse is a user-centric next-generation Internet with infinite possibilities. 
	
Wonderful visions in Metaverse are becoming reality with advances in computer and communications technologies. To name a few, advances in extended reality, e.g., virtual reality (VR) and augmented reality (AR), make it easy to access Metaverse~\cite{tang2022roadmap}. The development of hardware devices such as head mounted display (HMD) can solve the discomfort caused by VR, e.g., motion sickness. Second, developments in computational graphics make the customisability of virtual services feasible, and can render high-quality virtual objects to users. Third, significant improvements in wireless communications allow us to make full use of network resources to support the massive data transmission required by Metaverse services. With effective network resource allocation algorithms and promising sixth-generation (6G) communications, the Quality-of-Experience (QoE) can be significantly enhanced.
	
Leveraging on technological advances combined with business potentials, large enterprise companies, e.g., Facebook, Epic Games, and Tencent, have announced their entry into Metaverse market. For example, the social media giant Facebook re-brands itself as ``Meta" and launches an Metaverse meeting application named {\textit{Horizon Workrooms}}. Moreover, the Seoul Metropolitan Government is going to release {\textit{Metaverse Seoul}} in Seoul Vision 2030. An insight is that most Metaverse applications are user-oriented. This leads us to think about how to design better user-centric customized Metaverse services to maximize the QoE.
	
	However, novel yet sophisticated Metaverse services require high-speed real-time data processing and high-rate data transmission, which places huge demands on network resources, e.g., virtual traveling, virtual meeting, and virtual game. Metaverse service providers (MSPs) need to ensure high standards of sensing, transmitting, and rendering, to bring users the natural feeling of immersion. Thus, to make full use of network resources, an intuitive idea is to allocate more resources to the virtual objects in Metaverse that are of more interest to users. In this paper, we define the term {\textit{``attention''}} to describe how much the user cares about the object, which is driven by the user's interest. The problem is that the historical user-object-attention value record is sparse, because we can only obtain user attention values for a small part of the virtual objects that they have seen. Specifically, considering the personalized and heterogeneous nature of the user attention mechanism, we have to answer the following two questions:
	\begin{itemize}
	\item {{How can we predict accurately the attention value of a user to any virtual object using historical sparse user-object-attention records?}}
	\item {{How to model the QoE mathematically using the user-object-attention values, and accordingly guide the design of network resource allocation algorithms?}}
	\end{itemize}
	As shown in Fig. \ref{Attenextra}, we discuss the related Metaverse techniques and datasets, and propose several feasible solutions to address the above two problems in this article. In general, our contributions can be summarized as follows:
	\begin{figure*}[t]
		\centering
		\includegraphics[width=0.97\textwidth]{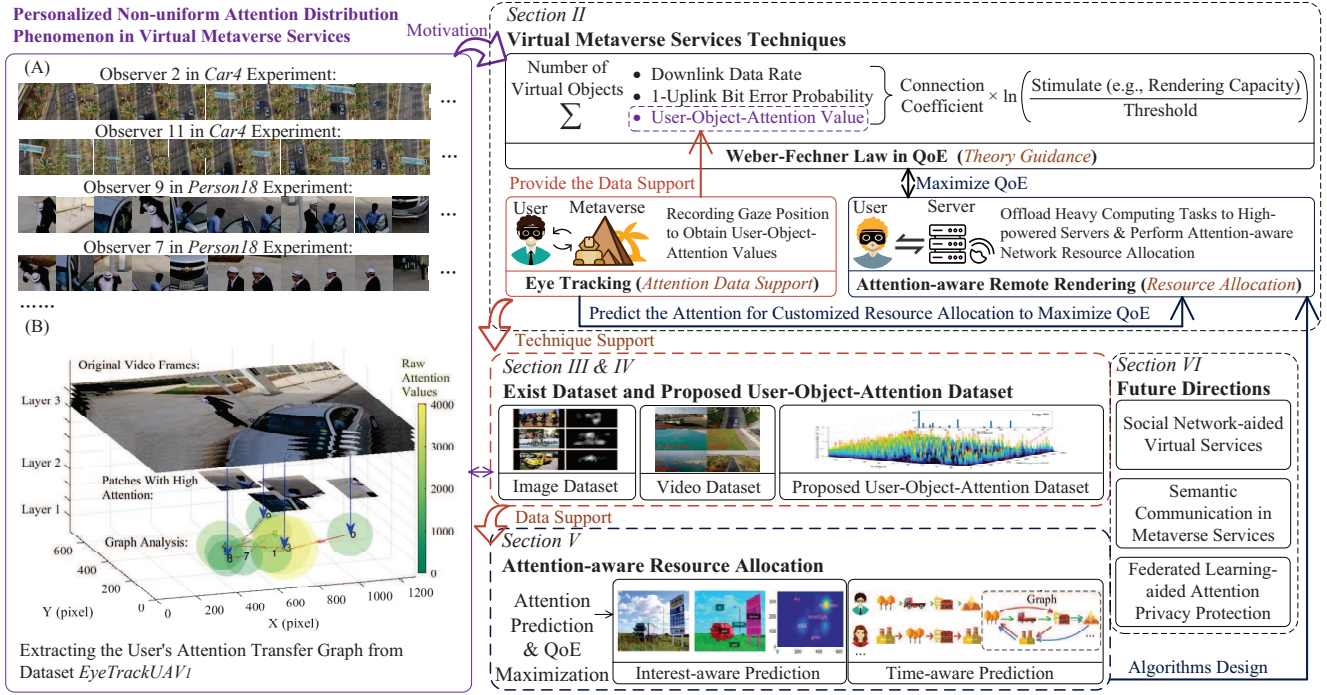}
		\caption{Structure of the article.}
		\label{Attenextra}
	\end{figure*}
\begin{itemize}
	\item We review three key technologies related to customized Metaverse services, and conduct a comprehensive discussion of available datasets. We then propose our dataset, i.e., User-Object-Attention Level (UOAL)\footnote{Github link of UOAL: https://github.com/HongyangDu/User-Object-Attention-Level.}, that other researchers can use to test and validate their new algorithms and services effectively. Different from existing datasets, UOAL provides a large number of ground truth personalize attention values of $30$ users to $1,000$ images of potential scenes in Metaverse. Specifically, practical sparse user-object-attention records can be simulated from UOAL, which can be used as the input in the design of attention prediction algorithms. Ground truth attention values in UOAL can be used as benchmarks to verify the accuracy of predictions.
	\item Based on a concise tutorial on how to use UOAL, we propose several feasible user attention prediction algorithms, which can be divided into two categories, i.e., interest-aware and time-aware prediction schemes. The former can be applied when the user's long-term attention history is available, and the latter can be used for attention prediction in a single service session without user identification. This provides feasible directions for future attention prediction algorithms design research.
	\item We discuss how to apply the Weber-Fechner's Law, a formula widely used in the field of psychology, to model the QoE in Metaverse. A generic form of QoE is proposed and can be adjusted according to the specific type of stimulus in different Metaverse services. With the help of the QoE metric and predicted user attention values, the network resource can be allocated optimally.
\end{itemize}
	
	\section{Customized Metaverse Services Techniques}
	In this section, we review the features of Metaverse services and three related techniques/principles to achieve customized service design, i.e., Weber-Fechner law in QoE, remote rendering, and eye-tracking. 
	
	\subsection{Features of Metaverse Services}
One of the main reasons for the popularity of Metaverse and web 3.0 concepts is that services are user-centric, which is also the fundamental basis of the next-generation Internet. In Metaverse services, each MSP needs to transmit a large amount of virtual object data to users, which requires a huge amount of network resources. However, from a psychological point of view, the quality of virtual objects to which users pay more attention can mainly affect the QoE perceived by users.

To take advantage of this fact to design attention-aware resource allocation algorithms, we introduce three related techniques/principles. Specifically, when a user chooses one type of Metaverse services, the Weber-Fechner law provides theoretical guidance for QoE modeling. Then, the eye-tracking technique in the user's HMD provides data support for user-object-attention values prediction. Remote rendering can perform network resource allocation algorithms based on predicted user-object-attention values.
	
	\subsection{Related Metaverse Services Techniques}

	\subsubsection{Weber-Fechner Law in QoE}\label{QoES} 
	
	The quality of service (QoS) attempts to measure objectively service performance metrics, i.e., bit error probability and data rate. In contrast to QoS, QoE is a subjective measure from the perspective of the user.
	A psychological theory that has a potential to be widely used in QoE analysis is the Weber-Fechner's law, which points out that the perception of the human sensory system is logarithmically related to the magnitude of the stimulus~\cite{reichl2010logarithmic}. The effectiveness of Weber-Fechner's law has been experimentally validated in a wide range of sensory perceptions, and has been applied to QoE assessment in communication networks~\cite{reichl2010logarithmic}.
	
	
	In Metaverse services, users can receive a variety of stimuli, such as visual stimuli given by rendered virtual objects, auditory stimuli given by sounds, and tactile stimuli brought by the tactile Internet of Things technique. In addition, since stimulation occurs in the virtual world, we need to consider the connection between the virtual and the real worlds. As shown in Fig.~\ref{Attenextra}, a workable mathematical form of the QoE can be expressed as a ${\textit{connection coefficient}}$ multiplied by the logarithm of ${\textit{stimulus intensity}}$. Considering that the downlink needs to transmit data quickly and the uplink needs to upload user interaction instructions accurately, we can multiply the ${\textit{downlink data rate}}$ by ${\textit{one minus the uplink bit error probability}}$ and then multiply the ${\textit{user-object-attention value}}$ to obtain the ${\textit{connection coefficient}}$. The ${\textit{stimulus intensity}}$ depends on the type of Metaverse services, e.g., for virtual services, it can be expressed in terms of the resolution of the virtual object. Note that other performance metrics, e.g., downlink bit error probability, can also affect QoE, but may not be main aspects. Fortunately, the ${\textit{connection coefficient}}$ in QoE formulation can be adjusted according to the practical Metaverse services.
	

	
		\subsubsection{Eye Tracking}
	Eye movement is the manifestation of human attention. In free-viewing, the duration of one's gaze on an object depends mainly on the level of interest. In general, it has been empirically proved that eye movement data can be used to evaluate the attention of users, detection of personality traits, and activity recognition even in challenging everyday tasks~\cite{braunagel2017online}. Additionally, the study of user-object-attention has provided novel ways for many scientific fields, such as data compression, repositioning, and decision-making.
	
	
	
	
	Recent advances in head-mounted display (HMD), embedded cameras and computer graphics make it easier to obtain eye movement data. Specifically, a fixation point detection method for HMD eye trackers is proposed in \cite{steil2018fixation}. This method can obtain information from small areas around fixation points, and analyze the appearance similarity to detect fixation points. Moreover, to reduce user concerns about privacy, eye movement data protection techniques were proposed. Through a mechanical shutter, a design named PrivacEye \cite{steil2019privaceye} can automatically enable and disable the first-person camera of the eye tracker. We can observe that the eye movement data collection technologies are continuously improving and the relevant datasets are increasing, which provides strong support for further research on user attention. Considering that VR and AR are currently the dominant accesses to Metaverse, it is feasible to use eye-tracking data to help network resource allocation in customized Metaverse services. 
	
	
	

	\subsubsection{Attention-Aware Remote Rendering}
	To create VR visuals, traditional VR implements all rendering commands at full resolution to provide stereo view pairs for both left and right eyes. This not only doubles the workload, but also introduces additional overhead when switching between render states. To address this limitation, several solutions based on the graphics processing unit (GPU) have been proposed~\cite{hubner2006multi}. 
	However, HMD devices might become overweight and overheat due to the new rendering hardware deployment. To solve this problem, remote rendering is used, which enables high-quality VR on low-powered devices. This is achieved by offloading the heavy computing and rendering to high-powered servers that stream VR as video to the client side. In addition, when the rendering task is performed on the server side, customized and perception-based rendering can be performed according to the user's attention. To further improve the user experience and reduce compute power, we propose an attention-aware rendering scheme in Section~\ref{S4A}, where the computing resources are allocated to areas that have a higher impact on Metaverse users' perception.

	\section{Dataset for Metaverse Research}\label{s3}
%
	In this section, we investigate some eye-tracking datasets and then present the UOAL to provide high-quality data support for the design of customized Metaverse services.
	\subsection{Existing Datasets}
	\subsubsection{Image Datasets}
	
	A series of datasets of free-viewing eye-movement recordings are presented in \cite{wilming2017extensive}, which contain fixation locations from 949 observers on more than 1000 images. All studies allow free eye movements, and differ in the participants' ages, stimulus sizes, stimulus modifications, and stimuli categories. However, the datasets in \cite{wilming2017extensive} focus on predicting a universal saliency map across all observers, without considering that the visual attention of different observers varies significantly under certain circumstances. To solve this problem, a personalized saliency detection dataset named PSM is presented in \cite{xu2018personalized}. Although PSM has sufficient data, it is not proposed for Metaverse research and therefore the images and labels cannot be used in a straightforward manner.

	
	\subsubsection{Videos Dataset}
	Video datasets are equally significant as images. Two datasets of human visual behavior during the observation of unmanned aerial vehicles (UAVs) videos, i.e., EyeTrackUAV1 and EyeTrackUAV2, were presented in \cite{perrin2020eyetrackuav2}. They consist of a collection of precise binocular gaze information over 43 videos. Thirty participants observed stimuli under both free viewing and task conditions. Video databases can provide more information than image databases, but the number of observers in existing video databases is usually small.
	
	
	\subsubsection{Insights}
	From the above datasets, we can observe that the user's attention is dispersed to images or videos. For example, in Fig. \ref{Attenextra}, we can see that the high attention patches of different observers are different in the same video: In the video segment {\it Car4}, the observer 2 cares more about the road while the observer 11 cares more about the road signage. In the video segment {\it Person18}, the observer 9 cares more about the passenger while the observer 7 is more attracted to the driver. When users experience Metaverse services using VR or AR, they will also have different attention values to different virtual objects. Thus, attention-aware resource allocation algorithms can be designed to achieve high-quality customized Metaverse services.
	However, none of the datasets aforementioned is specifically proposed for guiding customized power allocation schemes in Metaverse. Specifically, existing datasets have several shortcomings for Metaverse research:
	\begin{itemize}
	\item Not all images are suitable for simulating customized Metaverse services, such as meaningless geometric images and movie scenes.
	\item Although there are a large number of images and observers involved in the existing datasets, the observers did not see all the images. Instead, they were grouped to check different images.
	\item The segmentation labels in many datasets such as PSM are all manually annotated, which leads to one object having different labels.
	\end{itemize}
	To overcome the shortcomings of existing datasets, we propose UOAL which will be universally useful in Metaverse research.
	\subsection{User-Object-Attention Level Dataset}\label{S4}
	To study the unevenness of user attention, we need to record the gaze positions of multiple users on multiple objects. We then introduce the creation process of UOAL in detail, and provide a concise tutorial on how to use UOAL.
	\begin{figure*}[t]
	\centering
	\includegraphics[width=1\textwidth]{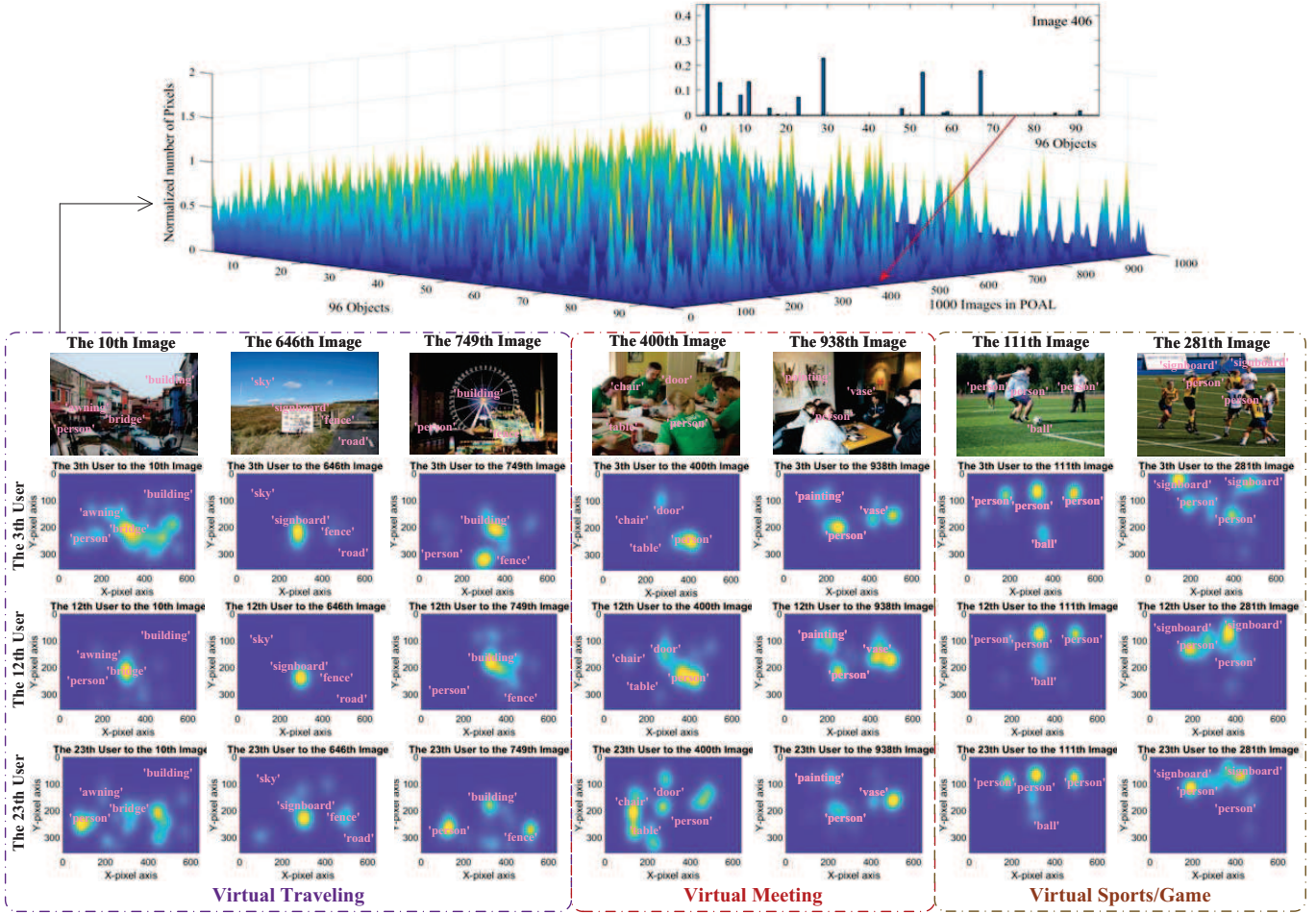}
	\caption{Some examples in our proposed user-object-attention level dataset for different types of customized Metaverse services research.}
	\label{UOAL}
	\end{figure*}
	
	\subsubsection{Image Dataset Pre-processing}
	We use the phenomenon of uneven attention distribution of users when viewing images to simulate the different attention of users to virtual objects in Metaverse. We build the UOAL mainly based on the PSM dataset~\cite{xu2018personalized} that contains the attention values of $30$ observers to $1,600$ images. From the PSM, we manually select $1,000$ images and corresponding eye-movement recording data. The selected images are related to Metaverse scenes to overcome the limitations in PSM. Then we re-scale all selected images to $360$ $\times$ $640$ pixels.
	
	\subsubsection{Object Segmentation Labels Acquisition}
	To avoid errors caused by manual annotations to objects, e.g., different naming of the same object in PSM, we use K-Net~\cite{zhang2021k}, a state-of-the-art semantic segmentation algorithm, to determine object labels and pixel positions of each object. After semantic segmentation, we obtain $150$ object categories. By filtering, we obtain $96$ high-quality object labels.
	
	\subsubsection{Scenes Grouping}\label{sec:group}
	Typically, MSPs can provide users with several different styles of virtual service options. Each user may have experienced a different number of services. To simulate this, we split the $1,000$ images in UOAL into $5$ same-style groups using the $K$-nearest neighbors (KNN) algorithm. By calculating the number of pixels occupied by each object in the segmentation labels, we generate a normalized vector for every image with a length of $96$, in which each element reflects the users' attention to the corresponding object, as shown in the top of Fig.~\ref{UOAL}.
	
	\subsubsection{Attention Values}\label{sec:process}
	The attention value represents the user's interest in the object. We use segmentation labels and eye movement records to quantify the attention value of each user to each object in the UOAL. Specifically, for one user, we consider that the segmentation label of each object has a corresponding {\textit{pixels vector}} whose length is the number of pixels in the label. The element in the vector is the corresponding attention value~\cite{xu2018personalized}, which is shown in Fig.~\ref{UOAL} as the heat-map. Then, we can denote all labels appearing in each image by an {\textit{objects vector}} whose length is the number of objects in the image. The element in the {\textit{objects vector}} is the sum of elements in {\textit{pixel vector}} of the corresponding object. Thus, the user's attention to the object, e.g., {\textit{object A}}, can be formulated as the sum of the pixel vector elements of {\textit{object A}} divided by the sum of all elements of the vector where {\textit{object A}} appears. For example, {\textit{object A}} appears in $3$ images that a user has seen. The numbers of pixels of {\textit{object A}} are $100$, $300$, and $200$, respectively. The gaze records of the user to the corresponding positions are calculated as $20$, $30$, and $40$, respectively. Thus, the attention value of the user to {\textit{object A}} is $(20+30+40) / (100+300+200) $.
	Motivated by the design of a commonly used psychometric scale named $5$-Point Likert scale~\cite{joshi2015likert}, we split the users' attention values to the $96$ objects into $5$ attention levels, i.e., $1,\ldots,5$.

	\subsubsection{Customized Metaverse Service Simulation}
	Note that the user-object-attention values calculation method discussed in Section~\ref{sec:process} is based on the fact that the user participates in all virtual services. Thus, the obtained results are the ground truth that reflects the real attention of users.
	
	However, to simulate practical user-object-attention records, we need to consider that the users might only have participated in a few Metaverse services, and users have only seen parts of the images according to the paths that they freely choose in the virtual world. Thus, we next discuss how to use our UOAL dataset to simulate sparse user-object-attention records, to provide input to the attention prediction algorithms of researchers.
	
	Recall that we split images into $5$ groups in Section \ref{sec:group}, which can be used to simulate $5$ virtual service options. The options and number of services that users selected should be random. Furthermore, after a user selected a virtual service option, the images in UOAL that the user has seen should be random because that different traveling paths may be chosen. Therefore, we can re-perform the process in Section~\ref{sec:process}, and obtain user-object-attention records with lots of empty level values. The process of obtaining sparse attention records is given in Tutorial \ref{algorithm1}. With the obtained sparse records as input, the attention prediction methods are presented in Section~\ref{feaf}.
	
	\floatname{algorithm}{Tutorial}
	\begin{algorithm}[t]
	\caption{Generating sparse historical user-object-attention records for the user in customized Metaverse service.} 
	\label{algorithm1}
	\hspace*{0.02in} ${\bf{Input:}}$
	Five grouped images, user-object-attention records, and segmentation object labels from UOAL.\\
	\hspace*{0.02in} ${\bf{Output:}}$
	Sparse user-object-attention values for the user.
	\begin{algorithmic}[1]
	\State Generate a random number ${\textit{ran}}_1$ from $\left[2,4\right]$, which indicates that the user has participated in ${\textit{ran}}_1$ different kinds of services.
	\State Generate a random number ${\textit{ran}}_2$ from $\left[30,70\right]$ to simulate that users might walk different routes in virtual traveling.
	\State Randomly select ${\textit{ran}}_1$ image groups and reserve ${\textit{ran}}_2\%$ images.
	\State{Initialize attention value array for each object in the selected segmentation labels}.
	\For{Every virtual object in selected labels}
	\State{Count the number of occurrences of the object}.
	\State Find the position (pixels) of the object.
	\State Calculate the attention value of the object: Sum the user-object-attention values at position identified in Step 7, and divided by the frequency calculated in Step 6.
	\EndFor
	\State Map attention values into $5$ levels to obtain sparse user-object-attention values.
	\end{algorithmic}
	\end{algorithm}

	\section{Attention-Aware Resource Allocation}\label{S4A}
	The development of Metaverse and web 3.0 has brought a rich service experience to users. To provide users with customized services, we can predict accurately users' attention, so as to devote network resources according to users' interests. Using the historical sparse attention records of users, e.g., from UOAL using Tutorial \ref{algorithm1}, we design the attention-aware resource allocation scheme that includes two steps: 
	\begin{itemize}
		\item Attention Prediction: After the user selects a virtual service option, the MSP predicts the user's attention values to all virtual objects.
		\item QoE Maximization: The MSP maximizes the QoE to obtain an optimal network resource allocation scheme.
	\end{itemize}
	
	
	\subsection{Step 1: Attention Prediction}\label{feaf}
In this Step 1, we discuss how MSPs can apply historical data to predict unknown attention values. We divide the prediction methods into the interest-aware and time-aware prediction, depending on whether the user's long-term historical record is available.
	\subsubsection{Interest-Aware Prediction}
	\begin{figure}[t]
	\centering
	\includegraphics[width=0.4\textwidth]{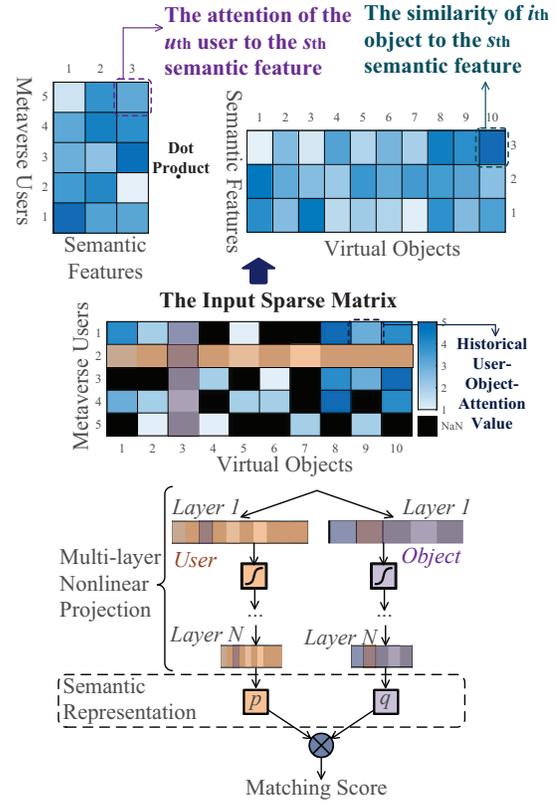}
	\caption{The methods for interest-aware attention prediction.}
	\label{MF1}
	\end{figure}
	For users who access Metaverse using the same identifier each time, the MSP can obtain long-time historical attention records. As a result, the MSP can predict users' intrinsic interests, i.e., attention to virtual objects. To achieve interest-aware prediction, we propose the following two approaches:
	
 \paragraph{Content-based approach}
	MSPs can create a profile for each virtual object or user in Metaverse to characterize its properties. For example, an object's profile in a virtual traveling service may include attributes such as type and popularity. A similar example that has succeeded in the real world is the Music Genome Project~\cite{joyce2006pandora}, in which a trained music analyst rates each song based on hundreds of musical characteristics. User profiles can be collected through questionnaires when joining Metaverse. Content-based approach infers the user attention by analyzing and comparing the profiles of users and objects. However, this approach requires a lot of user information, some of which is unavailable or difficult to obtain.
	
\paragraph{Semantic-aware approach}
	An alternative approach is to analyze the user's historical attention records from implicit feedback, such as the duration of gaze, without requiring the creation of explicit profiles. Although the user-object-attention records in Metaverse are typically sparse, we can mine users' attention through semantic level analysis. The semantic-aware approach describes objects and users by semantic vectors inferred from the obtained historical records. Specifically, we propose the following two methods:
	\begin{itemize}
	\item $\textit{Matrix factorization-based method:}$ Both users and objects are mapped to a joint semantic space with dimension of $f$. For a given virtual object, the semantic vector measures the degree to which the virtual object possesses $f$ features. For a given Metaverse user, the semantic vector measures the attention level to the corresponding feature. Thus, the sparse user-object-attention values are modeled as inner products in the semantic space, as shown in the upper half of Fig.~\ref{MF1}. Note that the dimension of the semantic vector could be interpretable, e.g., a {\it{football}} could be $90\%$ of a {\it{toy}} and $30\%$ of a {\it{tool}}, or completely uninterpretable, e.g., several deep-seated factors that influence a person's personality. The major challenge is computing the mapping of each object and user to factor vectors. One conventional method is singular value decomposition (SVD) commonly used to identify latent semantic factors in information retrieval.
	\item $\textit{Deep matrix factorization-based method:}$ The matrix factorization-based method performs one-layer linear projection to obtain the semantic vectors of users and objects. Given remarkable ability of deep learning methods in representing objects and users, we can use the neural network architecture to learn low dimensional semantic vectors to represent the users and objects. Inspired by the deep structured semantic models which have been proved to be useful for web search, recommendation system, and query-document matching, we consider an architecture of deep neural network to project Metaverse users and objects into a semantic space, as shown in the lower half of Fig.~\ref{MF1}.
	In our architecture, we have two multi-layer networks to transform the representations of user and objects vectors, respectively. Through the neural network, the user and object are finally mapped to low-dimensional vectors in the semantic space. The similarity between the input user and object is then measured by cosine similarity~\cite{yang2022semantic}.
	
	\end{itemize}


	\subsubsection{Time-Aware Prediction}
	\begin{figure}[t]
	\centering
	\includegraphics[width=0.45\textwidth]{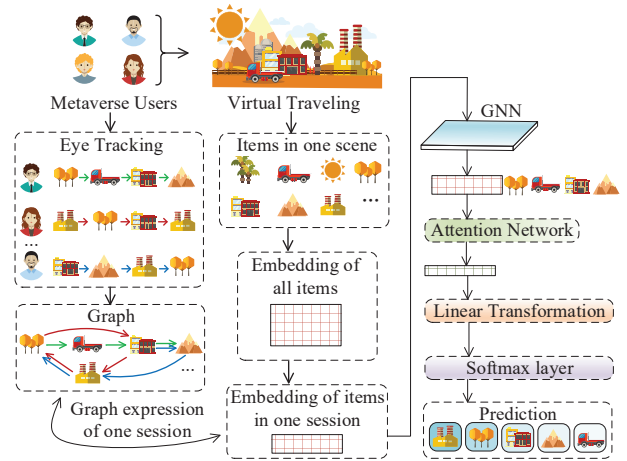}
	\caption{GNN approach for time-aware attention prediction in virtual traveling Metaverse service.}
	\label{AlgorithmModel}
	\end{figure}

In several Metaverse services, information about users may be unknown, and only user-object-attention values in the current ongoing service is available. Therefore, it is significant to model limited behaviors within a service session and generate accurate time-aware predictions accordingly \cite{li2022contextual}.
	
As shown in Fig.~\ref{AlgorithmModel}, Graph Neural Networks (GNN) can capture the transfer of user attention between objects and generate accurate embedding vectors accordingly, which is difficult to achieve by traditional sequential methods \cite{guo2020deep}. Taking the virtual traveling in Metaverse as an example, we can build the session graph based on the movement of the user's gaze point among virtual objects. Then, we process each session graph and obtain node vectors through the gate-control GNN. After that, we use the attention network to express each session as a combination of global attention and current preferences for that session. Finally, we predict the probability that each object is the next one to be watched for each session.


	\subsection{Step 2: Quality of Experience Maximization}
	In this Step 2, we maximize the QoE of users. Moreover, we compare QoE under attention-aware and uniform resource allocation algorithms.
	
Recall the mathematical form of QoE that we discuss in Section \ref{QoES} as an $N$-term summation expression, where $N$ is the number of virtual objects. Note that network resources such as transmit power, bandwidth, and computing resources all affect the value of QoE. The reason is that different allocation schemes result in different data rate, bit error probability and rendering quality. The constraints of the QoE maximization problem are that there are upper limits on transmit power, bandwidth, and rendering capacity. Fortunately, after Step 1, we can estimate the user attention value for each object. Then, the maximization problem can be solved by alternating optimization and convex optimization method.
	
	\begin{figure}[t]
	\centering
	\includegraphics[width=0.45\textwidth]{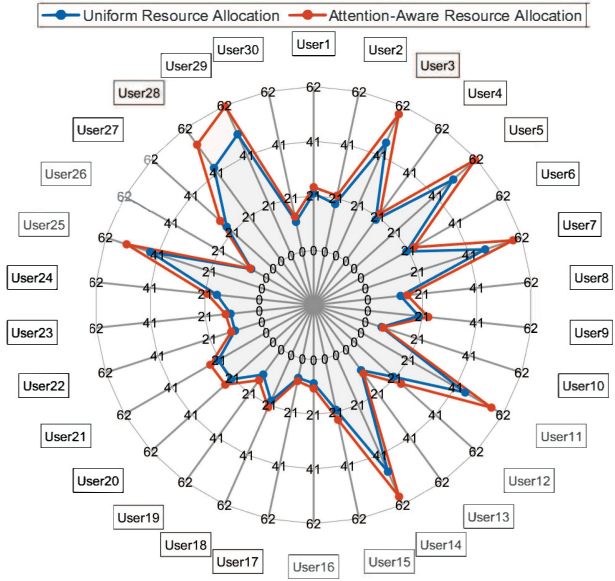}
	\caption{The QoE of 30 Metaverse users, with uniform and attention-aware rendering capacity allocation schemes.}
	\label{radar}
	\end{figure}
	Without losing generality, we take the allocation of rendering capacity as an example of network resource. Then we compare attention-aware and uniform rendering capacity allocation schemes in the stimulated VR setting. Figure~\ref{radar} illustrates the QoE of $30$ users under both allocation schemes. The resolution {\small $\left( {\rm K}\right)$} is used as the measure of rendering capacity. Here we define that 1 ${\rm K}$ resolution refers to $960\times 480$ pixel resolution. We consider that the minimal rendering capacity threshold for one virtual object is $15$ {\small ${\rm K}$}. The total rendering capacity for the $i^{\rm th}$ user is ${\textit{the number of virtual objects}}\times 20$ {\small ${\rm K}$}. According to Tutorial~\ref{algorithm1}, we simulate that the Metaverse service options are selected randomly by users. Then, different allocation schemes are used. Specifically, the uniform rendering capacity allocation scheme distributes uniformly the rendering capacity while ensuring that each object has a minimum threshold assigned. For the attention-aware scheme, the MSP predicts the user-object-attention values as we discuss in Section~\ref{S4A}. Thus, the rendering capacity is allocated to maximize QoE. From Fig.~\ref{radar}, we can observe that the attention-aware rendering capacity allocation scheme can achieve a maximum of $25.5\%$, a minimum of $6.26\%$, and an average of  $20.1\%$ QoE improvement than the uniform rendering capacity allocation scheme, in the simulations with $30$ users.
	
	\begin{figure}[t]
	\centering
	\includegraphics[width=0.45\textwidth]{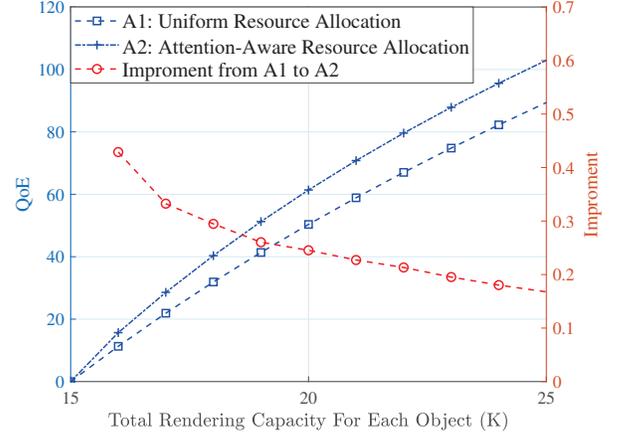}
	\caption{The QoE of one Metaverse user versus the average rendering capacity for each object. We consider that the numbers of data transmitting and receiving antennas are $6$ and $7$, respectively. There are $3$ paths of co-channel interference with $1$ ${\rm{dBW}}$. The distance between the BS and the RS is $10$ ${\rm m}$, and the path loss exponent is $2$.}
	\label{zhexian}
	\end{figure}
	Then we focus on the subjective experience of Metaverse user, i.e., the third user in Fig.~\ref{radar}. For the virtual scenario option selected randomly by the third user, there are $56$ objects. Figure \ref{zhexian} plots the QoE of the third user versus the total rendering capacity/{\textit{the number of virtual objects}} under three different resource allocation algorithms. An interesting insight is that the attention-aware scheme brings a higher percentage of improvement compared to the uniform allocation scheme when the total resources are constrained, i.e., when the total rendering capacity is small. This shows that the attention-aware scheme can improve resource utilization efficiency and thus bring a better Metaverse experience to users.

	\section{Future Directions}
	In this section, we discuss several research directions.
	
	\subsection{Social Network-Aided Virtual Services}
	Social network can be regarded as a graph with two core components, i.e., users and relationships. One advantage of using social networks to assist service design is that the cold start problem can be solved. For example, when a Metaverse user associates with a social networking account, such as Facebook, we are able to obtain the friend list of the user from Facebook. To some extent, the user's preferences can be inferred according to those of his/her close friends. However, representation and quantification of various relationship features remain challenging and worthy of further study.
	
	\subsection{Semantic Communication in Metaverse Services}
	Metaverse increases the demand for massive data transmission and storage. An increase in data rate brought about by physical layer technologies has gradually failed to keep up with the explosive growth of data volume. Thus, semantic communication is proposed as a technique that extends the conventional Shannon paradigm, and can reduce the data that needs to be transmitted and processed. One interesting research direction is that using semantic communication to achieve efficient data processing and storage involved in customized Metaverse services.
	
	\subsection{Federated Learning-Aided Attention Privacy Protection}
	Protecting the privacy of data relating to the user's attention, such as gaze position, is critical. Especially with the development of HMD devices, it has become easier to collect data. However, biometrics may be used for illegal purposes. As a distributed machine learning framework, federated learning (FL) can help MSPs to use attention data while meeting the requirements of privacy protection. However, the devices involved in FL, i.e., Metaverse accessing devices, are heterogeneous, and have different computing capabilities. To ensure the high-quality of the trained model, we need to develop the FL framework according to Metaverse characteristics.

	
	\section{Conclusion}
	In this article, we studied customized Metaverse services, aiming to make efficient use of network resources and improve QoE perceived by users. By predicting the user's attention values towards virtual objects, we can allocate more resources to objects in which users are more interested. We provided an overview of the key techniques required to design customized services. Moreover, a user-object-attention dataset, i.e., UOAL, was presented for Metaverse research. Then, we propose a two-step attention-aware network resource allocation algorithm. A tutorial overview was provided for two types of attention prediction methods, i.e., interest-aware and time-aware prediction. With the predicted attention values, QoE can then be maximized by allocating network resources optimally. Furthermore, we discussed open issues that are worthy of further research. It is hoped that this article will provide guidance regarding the consideration of user attention mechanisms in customized Metaverse services, to promote the construction of a user-centric next-generation Internet.

		\newpage
	\bibliographystyle{IEEEtran}
	\bibliography{IEEEabrv,Ref}
	\end{document}